\documentclass[%
 reprint,
 amsmath,amssymb,
 aps,
]{revtex4-2}

\usepackage{graphicx}
\usepackage{dcolumn}
\usepackage{bm}
\usepackage{braket}
\usepackage{comment}
\usepackage{hyperref}
\usepackage{verbatim}
\usepackage[nameinlink,poorman]{cleveref}
\usepackage{xcolor}

\crefname{equation}{Eq.}{Eqs.}
\crefname{figure}{Fig.}{Figs.}
\crefname{table}{Table}{Tables} 
\crefname{section}{Section}{Sections}
\crefname{chapter}{Chapter}{Chapters}
\crefname{appendix}{Appendix}{Appendices}
\crefname{algorithm}{Algorithm}{Algorithms}
\crefname{theorem}{Theorem}{Theorems}
\crefname{defn}{Definition}{Definitions}
\crefname{definition}{Definition}{Definitions}
\crefname{azm}{Assumption}{Assumptions}
\crefname{corollary}{Corollary}{Corollaries}
\crefname{lemma}{Lemma}{Lemmas}
\crefname{thmprop}{property}{properties}
\crefname{proposition}{Proposition}{Propositions}
\crefname{remark}{Remark}{Remarks}

\begin{document}

\preprint{APS/123-QED}

\title{Improved quantum sampling methods for molecular simulations}

\author{Connor van Rossum}
\affiliation{School of Mathematics and Physics, University of Queensland,  Qld 4072, Australia
}%
\thanks{These authors contributed equally to this work.}
\email{c.vanrossum@uq.edu.au}

\author{Jeffery Cohn}%
\affiliation{The MITRE Corporation, 200 Burlington Road, Bedford, MA 01730-1407 USA}%
\thanks{These authors contributed equally to this work.}
\email{jrcohn625@gmail.com}

\author{Sally Shrapnel}
\affiliation{School of Mathematics and Physics, University of Queensland,  Qld 4072, Australia \relax
}%

\author{Riddhi S. Gupta}

\affiliation{Joint School of Electrical Engineering and Computer Science (EECS) and the School of Mathematics and Physics (SMP), University of Queensland,  Qld 4072, Australia
}%

\date{\today}

\begin{abstract}
Quantum-selected configuration interaction (QSCI) methods use a quantum computer to identify dominant electronic configurations in the molecular ground state, while a classical computer diagonalizes the Hamiltonian within the reduced subspace spanned by those configurations. Sample-based quantum diagonalization (SQD), a leading QSCI approach, uses iterative classical post-processing to correct noisy quantum measurement to ensure that the corresponding configurations remain physically sensible. In this work, we show that SQD performance can be strongly influenced by uncontrolled growth of the classical diagonalization subspace. When classical resources are not explicitly constrained, classical uniform random sampling can reproduce SQD benchmarks as noise increases the diversity of sampled configurations. We show any fair benchmarking protocol of SQD must explicitly control diagonalization size over unique samples. We then address the problem of efficiently discovering physically relevant, energy-lowering configurations by introducing a measurement protocol based on non-orthogonal configuration interaction (NOCI). By distributing measurements across orbital bases optimized with respect to the molecular Hamiltonian, we obtain improved sample efficiency relative to measurements performed solely in the Hartree--Fock basis. Importantly, these improvements persist even under fixed classical resource budgets, demonstrating that the resulting configurations are of higher quality rather than being more numerous. Under our proposed benchmarking procedure, we establish measurement-basis engineering as a promising route to improving quantum sampling methods for electronic structure.
\end{abstract}

\maketitle


\section{\label{sec:intro}Introduction}

Simulating the electronic structure of molecules using a quantum computer is a problem of long-standing interest \cite{Aspuru-Guzik2006Apr,Lanyon2010Feb}. In a fault-tolerant regime, the foundational quantum algorithm for extracting molecular energies from simulations of molecular Hamiltonians relies on phase estimation \cite{Whitfield2011Mar}. These approaches require input quantum states that have a good overlap with some true, target ground state. Variational quantum eigensolvers (VQEs) \cite{peruzzoVariationalEigenvalueSolver2014, cerezoVariationalQuantumAlgorithms2021} and quantum-selected configuration interaction (QSCI) \cite{Kanno2026Jun,Robledo-Moreno2025Jun} have both emerged as hybrid quantum-classical sub-routines to prepare quantum states that provide this desired overlap. Of independent interest has been the capacity of these methods to compute ground state energies of molecular Hamiltonians. Variational quantum methods are guaranteed to converge to ground state energies from above \cite{szabo_modern_1996, peruzzoVariationalEigenvalueSolver2014} but suffer from scalability challenges \cite{cerezoVariationalQuantumAlgorithms2021, mccleanBarrenPlateausQuantum2018} as system size increases. One way to address this scalability challenge has been to split the VQE task of variationally discovering good ans\"atze quantum states from the task of diagonalization, leading to the development of QSCI methods. In QSCI, a quantum computer is used only to select candidate basis vectors (configurations) while diagonalization is carried out classically \cite{Kanno2026Jun,Robledo-Moreno2025Jun}.

Among existing QSCI approaches, sample-based quantum diagonalization (SQD)\cite{Robledo-Moreno2025Jun} has emerged as a leading method to run QSCI protocols on realistic quantum hardware, and has spawned several variants \cite{yuQuantumCentricAlgorithmSampleBased2025b, shajanQuantumcentricSimulationsExtended2024b, danilovEnhancingAccuracyEfficiency2025b, barisonQuantumcentricComputationMolecular2025}. Designed to accommodate \textit{noisy} quantum sampling where physically-motivated constraints of the underlying chemical system are inadvertently violated, SQD combines quantum sampling with iterative classical post-processing that probabilistically corrects configurations sampled on a noisy quantum computer prior to performing classical diagonalization. The method has demonstrated promising theoretical and experimental results on large molecules and standard thermochemistry benchmarks \cite{Raisuddin2025Nov}. 

Most impressively, SQD has been used to characterize fragments of a 300-atom mini-protein \cite{Shajan2026Jun} and a 12,000-atom protein-ligand complexes \cite{MerzKenneth2026May}. Large scale demonstrations such as these, have been enabled through embedding and fragmentation approaches that reduce the effective active-space size prior to quantum sampling and diagonalization. Recent developments include methods that combine SQD with fragment-based multireference theory to efficiently approximate strongly correlated active spaces \cite{wangLocalizedSamplebasedQuantum2026}.

Despite recent progress, concerns have emerged that these quantum methods may suffer from sampling inefficiencies that may preclude any computational advantage \cite{Reinholdt2025Jul}. Understanding sampling concentration challenges of QSCI more broadly is essential for assessing and benchmarking performance of specific demonstrations such as SQD. At the heart of these quantum sampling methods is the idea that repeated measurements of a quantum state preferentially sample the most important electronic configurations. This sampling procedure provides a compact set of basis vectors on which the molecular Hamiltonian can be classically diagonalized \cite{Kanno2026Jun,Robledo-Moreno2025Jun}. By configuration, we mean a particular assignment of the $N_e$ electrons to the $\kappa$ available spin-orbitals and each configuration corresponds to a Slater determinant. 

The success of quantum sampling therefore depends heavily on whether a final quantum state yields probability amplitudes that avoid a potential sampling bottleneck. Namely, for a highly concentrated probability distribution over electronic configurations, repeated measurements predominantly reproduce the same few dominant configurations, making the discovery of the remaining relevant configurations increasingly expensive. Conversely, broader distributions generate larger numbers of unique configurations, but may devote significant sampling weight to configurations that contribute little to lowering the ground state energy, since the identities of the most important configurations are generally not known \textit{a priori}. In the extreme case, such distributions can approach uniform sampling over the configuration space. The ideal target would be a state whose Born distribution is uniform over a fixed number of the most energy-lowering configurations. However, preparing such a state requires \textit{a priori} knowledge of the very configurations one seeks to discover.

\begin{figure}[!t]
    \centering
    \includegraphics[width=1.0\linewidth]{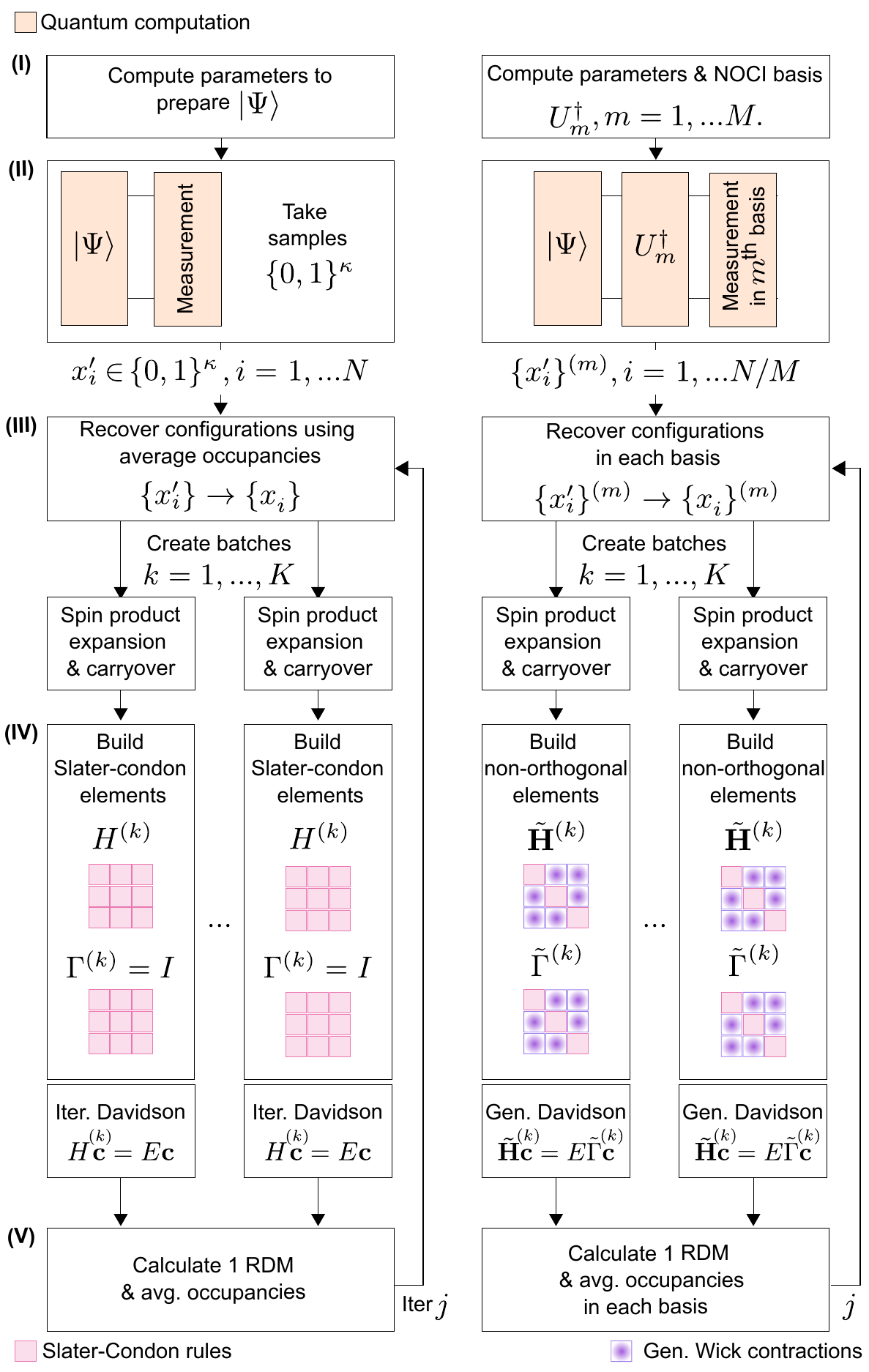}
    \caption{Low-energy states of  a molecule are associated with an initial quantum state $|\Psi\rangle$ \textbf{(I)} measured $N$ times in the standard computational basis \cite{Robledo-Moreno2025Jun} (left) vs. non-orthogonal bases (right). \textbf{(II)} $N$ binary measurement outcomes $x_i, i=1,\hdots,N$ on $\kappa$-qubits are collected in the computational basis (left) or spread across $M$ non-orthogonal bases defined by $\{ U_m \}$. \textbf{(III)} During configuration recovery, any measured $x' \in \chi'$ with an incorrect particle number are probabilistically bit-flipped to physical configurations $x \in \chi$ based on average occupancies over the full ensemble. Batches, $K$,  of size $r_0 \ll N/M$ are formed by subsampling corrected measurements. Any carryover configurations are appended, to form batches of size $r_j$, and spin-product expansion is performed within each batch. \textbf{(IV)} For each $k$-th batch, one constructs the Hamiltonian $H^{(k)}$ ($\tilde{\mathbf{H}}^{(k)}$) and the overlap matrix $\Gamma$ ($\tilde{\Gamma}$) using Slater-Condon rules within each basis (generalized Wick contractions between non-orthogonal bases) and an eigenvalue equation is solved via an iterative (generalized) Davidson method. \textbf{(V)} The smallest eigenvalue/vector pair is identified across all $K$ subspaces, and along with the one-electron transition density matrix, it is used to calculate the average occupancies to inform the $j+1$ round of step \textbf{(III)}. See \cref{app:background} Table II for notation.}
    \label{fig:draftcartoon}
\end{figure}

In this work, we investigate these concentration challenges and whether they can be overcome by combining information from multiple measurement bases. As illustrated in \cref{fig:draftcartoon}, our approach (right) distributes measurements across multiple optimized orbital bases rather than measuring exclusively in the Hartree--Fock basis (left). Samples from different bases correspond to non-orthogonal Slater determinants rather than repeatedly sampling the same Slater determinants in a fixed basis. Consequently, the measurement budget is spread over several complementary representations of the wavefunction, reducing redundant sampling within any one basis. 

Using the workflow shown in \cref{fig:draftcartoon} (left), we first illustrate and characterize symptoms of sampling concentration on established molecular energy benchmarks in literature. Disconcertingly, we find that replacing quantum measurements with uniform random configurations can outperform quantum sampling with SQD for a fixed number of raw, noiseless measurements. This effect occurs when classical post-processing procedures, including configuration recovery and spin product expansion, allow the diagonalization subspace to grow substantially beyond the set of configurations directly obtained from quantum measurements. In this regime, we additionally show that increasing local quantum noise naively increases the discovery rate of new determinants. That is, \textit{adding noise} seemingly \textit{improves} SQD performance. Our simulations have significant bearing on the naive interpretation of hardware experiments in literature. We confirm that these noise-enhanced improvements disappear once the resource budget for classical postprocessing is explicitly controlled, revealing that any meaningful interpretation of SQD experiments requires one to report detailed computational procedures in \cref{fig:draftcartoon} steps (III) and (IV).

Having established a fair benchmarking methodology, we then address the underlying concentration problem itself in \cref{fig:draftcartoon} (right). We combine SQD with non-orthogonal configuration interaction (NOCI) methods ~\cite{lowdin_quantum_1955,lowdin_quantum_1955-1,Thom2009Sep,Burton2019May,burton_generalized_2021,sun_selected_2024}, enabling measurements and Hamiltonian diagonalization to be performed in a non-orthogonal basis. We show that measurements performed in optimized non-orthogonal bases can identify useful, energy-lowering configurations more efficiently than measurements performed solely in the computational basis. Empirically, this approach improves SQD performance in terms of both the number of raw measurements and the diagonalization subspace size, providing evidence that the concentration challenge can be alleviated. We find that the distribution of the overall number of raw measurements across non-orthogonal bases can markedly change the efficiency with which lower-energy solutions are discovered. While constructing a non-orthogonal measurement basis requires classical preprocessing, the NOCI basis rotations can be combined with the final layer of orbital rotations typically used in SQD ans\"atze, such as the local unitary cluster Jastrow (LUCJ) ans\"atze, and therefore require no additional quantum resources to implement.

While our multi-basis measurement strategy does not solve the fundamental challenge of preparing high-overlap quantum states, it provides a practical way to mitigate these limitations. For a high-quality initial ansatz, our results show a clear advantage in energy convergence for a fixed measurement budget. Crucially, the method is most effective when the underlying ansatz is weak, where it accelerates the convergence of states that would otherwise be undersampled. The main trade-off is an increased classical cost arising from the construction and diagonalization of dense Hamiltonian and overlap matrices in a non-orthogonal basis. We view this approach as a promising first step toward boosting SQD sampling efficiency, with substantial opportunities for future algorithmic refinements. We present and discuss our results in \cref{sec:results,sec:discussion} before concluding in \cref{sec:conclusion}, with simulation details provided in \cref{sec:methods}. 

\section{Results \label{sec:results}}

\begin{figure*}[!t]
    \centering\includegraphics[width=1\linewidth]{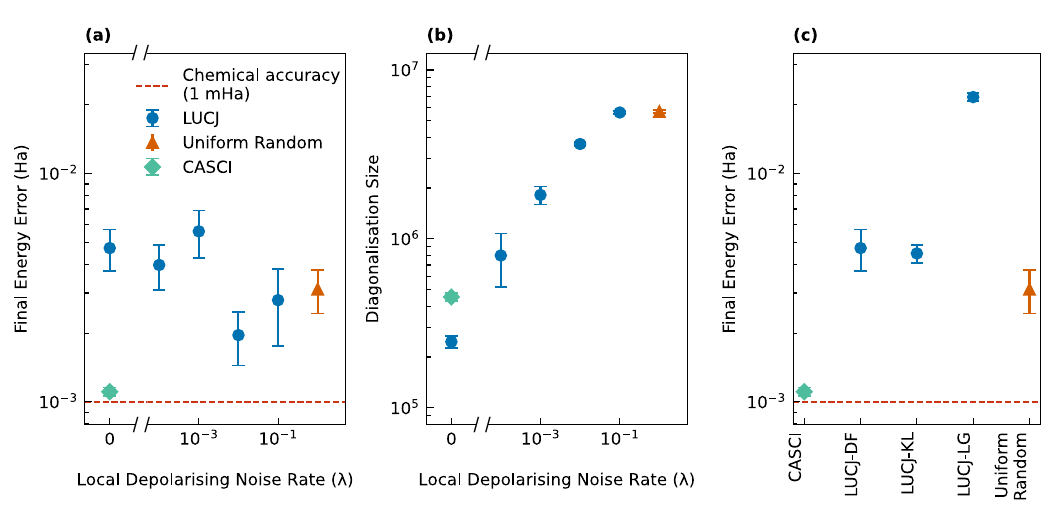}
    \caption{SQD for $\mathrm{N}_2$ in the 6-31g basis with two frozen orbitals and bond length $R=1.0\AA$. 
    (a) Final ground-state energy error for standard SQD under increasing local depolarizing noise rates applied to all two-qubit gates in the LUCJ circuit. The LUCJ ansatz (circles) is initialized via compressed double-factorization of CCSD $t_2$ amplitudes (LUCJ-DF) and optimized to minimize reconstruction error. This is compared to sampling from the exact CASCI wavefunction (“CASCI”, diamonds) or uniformly randomly from all configurations (triangles). Each data point uses $r_j = 2{,}668 \quad \forall j$ unique configurations for up-to five iterations, where only average orbital occupancies are propagated between iterations for configuration recovery. 
    (b) Growth of the diagonalization size $d$ for data in panel (a) due to Cartesian products of $\alpha$- and $\beta$-spin configurations. 
    (c) Comparison of alternative optimization strategies to LUCJ-DF, where LUCJ ans\"atze are optimized using KL divergence (LUCJ-KL) and log-likelihood (LUCJ-LG) with access to the exact ground state. Despite improved optimization, all LUCJ approaches remain comparable to uniform random sampling. In all cases, we set the carryover threshold to unity (see \cref{sec:methods}) so that each iteration does not inherit important samples discovered from previous iterations.
    }
    \label{fig:badresult}
\end{figure*}

\begin{figure}
    \centering
    \includegraphics[width=1.\linewidth]{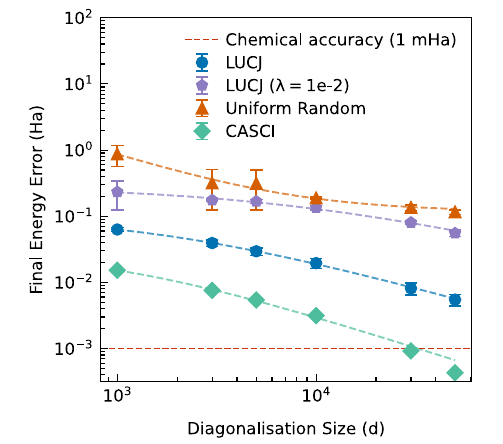}
    \caption{
    Benchmarking SQD for $\mathrm{N}_2$ in the 6-31g basis with two frozen orbitals and $R=1.0\AA$ under a fixed diagonalization size $d$, explicitly constraining classical computational resources. Subspaces are constructed directly from bitstrings sampled from measurement outcomes, without forming Cartesian products of $\alpha$- and $\beta$-spin configurations. The $x$-axis denotes the number of unique configurations used to define each diagonalization subspace ($r_j = d$). 
    For methods with noise, five iterations are performed, propagating only average orbital occupancies between iterations. 
    In contrast to \cref{fig:badresult}, controlling $d$ removes artificial improvements from subspace growth: energy error increases with noise and converges to uniform sampling from below.
    }
    \label{fig:good}
\end{figure}

Our results are presented in two parts. Firstly, we highlight the concentration issue by replicating existing benchmarks for $N_2$ molecules in Ref.~\cite{Robledo-Moreno2025Jun} using both the exact ground-state wavefunction as well as quantum ansatz states typically prepared on hardware. Here, sampling occurs by performing measurements in the Hartree--Fock computational basis. Secondly, we present results for sampling quantum states in the NOCI basis. Since NOCI methods are designed for strongly-correlated systems, we first present a stretched hydrogen chain, which serves as the primary benchmark used to characterize the method in the main text, with corresponding NOCI results for the $N_2$ molecule in supporting information.

To begin, we investigate how diagonalization subspaces grow substantially beyond the set of configurations directly obtained from quantum measurements by focusing on step (III) of \cref{fig:draftcartoon}. As detailed in \cref{sec:methods}, two dominant growth mechanisms involve spin-product expansion and carryover. In spin-product expansion, configurations associated with spin up/down are isolated and then recombined to increase the diagonalization subspace size either quadratically or by a constant factor in the number of raw samples. For carryover, important samples from $j-1$ round are automatically included in $j$-th round, such that diagonalization subspaces grow with each iteration. For simplicity, all our results ensure there is no carryover, hence minimizing the risk that the full Hilbert space is sampled for intermediate-sized active spaces.

\cref{fig:badresult}(a) replicates benchmarks in Ref.~\cite{Robledo-Moreno2025Jun} for the $N_2$ in the 6-31g basis with 2 frozen orbitals. We fix the subsampled batch size to $r_j=r_0=2,668$ for all iterations, corresponding to the average number of unique configurations in the support of the CASCI (Complete Active Space Configuration Interaction) wavefunction after $N=10^6$ measurements. The reference result obtained by sampling its CASCI Born distribution (`CASCI', green diamonds) is compared with uniform random sampling in the computational basis(`UR', orange triangles), which requires no quantum device. While this comparison faithfully reproduces the original result of Fig.~2 in Ref.~\cite{Robledo-Moreno2025Jun} where SQD applied to the exact wavefunction outperforms random sampling, we see that both noiseless and noisy LUCJ (blue circle) methods yield errors comparable to uniform sampling. In particular, increasing the local depolarizing noise rate $\lambda$ accelerates the discovery of low-energy configurations, resulting in reduced energy errors. In this regime, uniform random sampling replicates LUCJ based sampling SQD results.

This result of \cref{fig:badresult}(a) is better understood if one takes into account the actual growth in the spanning vectors used during diagonalization, $d$. In \cref{fig:badresult}(b), we plot $d$ corresponding to each data point in panel (a). The increase in $d$ with noise rate $\lambda$ arises from the Cartesian product of $\alpha$- and $\beta$-spin configurations, even when carryover between iterations is disabled, illustratively depicted in \cref{fig:draftcartoon} (III). Constructing subspaces via the Cartesian product of half-bitstrings is the default setting under which most SQD experiments in literature appear to have been performed using the software package in Ref.~\cite{qiskit-sqd}. Beyond increasing the classical resource cost for diagonalization, and implicitly sampling larger proportions of the full Hilbert space, using this subspace construction also destroys correlations that exist between spin-up and spin-down configurations. 

The extent to which a better LUCJ ansatz can outperform random sampling is addressed in \cref{fig:badresult}(c). The default SQD ansatz is an LUCJ state constructed using compressed double-factorization of $h_{pqrs}$ (`LUCJ-DF'), initialized from a CCSD calculation \cite{Robledo-Moreno2025Jun,qiskit-sqd}. This initialization may be further refined by optimizing the ansatz to minimize the reconstruction error of $t_2^{\mathrm{LUCJ}}$. This optimized LUCJ ans\"atze is compared with two alternative LUCJ optimization protocols, which both assume access to the exact ground-state solution: we minimize either the KL divergence (`LUCJ-KL') or the log-likelihood (`LUCJ-LG') between the Born distributions of the exact wavefunction and a candidate LUCJ state. Despite access to the exact ground-state wavefunction, all optimization strategies shown in \cref{fig:badresult}(c) exhibit the same trend observed in \cref{fig:badresult}(a), namely that noiseless LUCJ performs comparably to uniform random sampling.

Such simulations highlight that naive benchmarking of SQD on hardware could yield results that are entirely reproducible by uniform random sampling if the number of unique configurations $r_j$ and the size $d$ of the subspaces for classical diagonalization are allowed to grow. In addition to noise, the use of Cartesian products of $\alpha$ and $\beta$ spin configurations rapidly generates new unique configurations, largely independent of the underlying quantum state. Even for strongly correlated systems, such as a stretched $H_{8}$ hydrogen chain in the 6-31g basis, we find that uniform random sampling can outperform even sampling from the Full Configuration Interaction (FCI) solution after sufficient iterations of configuration recovery, attributed entirely to the differing sizes of $d$ for different sampling strategies (see \cref{app:ExtraFigs}). Any demonstrations that additionally use sub-unity carryover thresholds would further exacerbate this effect, whereby any inefficiencies in sampling quantum measurements are inadvertently hidden by classical post-processing.

As the diagonalization subspace dimension $d$ is the dominant computational resource in SQD, we now explicitly constrain classical computational resources by fixing $d$ in \cref{fig:good}. On the $x$-axis, we control the number of unique bitstrings used to construct each subspace and sample exclusively from measurement outcomes. That is, we forgo the Cartesian product of $\alpha$ and $\beta$ spin configurations. The final basis is constructed from symmetry-preserving bitstrings, with subsets of size $r_j = r_0 = d$ defining the diagonalization subspace. For uniform random sampling and noisy LUCJ, we perform five rounds of configuration recovery and diagonalization, consistent with \cref{fig:badresult}, where only the average orbital occupancies are propagated between iterations. In \cref{fig:good}, we see that ground-state energy error increases with local depolarizing noise in the LUCJ circuits (blue, purple), and converges to uniform random sampling from below if subspace sizes $d$ are controlled. We note that this simulation acts as a sense-check for how to benchmark SQD in noisy settings. Since the molecule and the size of the active space are identical in \cref{fig:badresult} and \cref{fig:good}, sampling from these LUCJ circuits would not be practically useful if they fail to outperform uniform random sampling for a computationally tractable choice of subspace dimension $d$. 

\begin{figure*}[!t]
    \centering
    \includegraphics[width=1\linewidth]{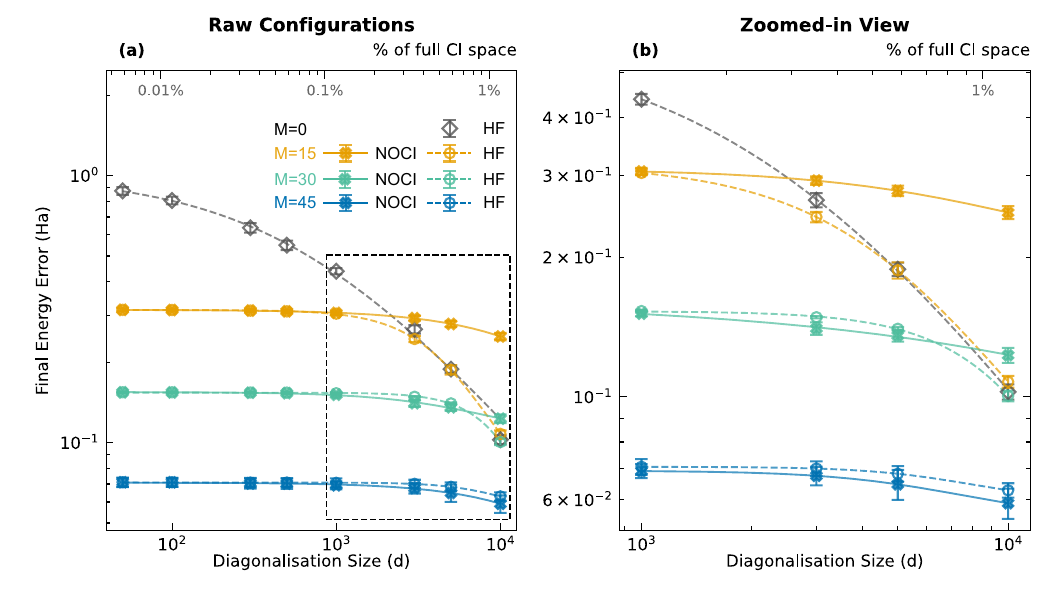}
    \caption{
    SQD using NOCI bases for $\mathrm{H}_{12}$ in the STO-6G basis with no frozen orbitals and $R=2.0\AA$ using a fixed diagonalization size $d$, without spin product expansion as in \cref{fig:good}. Samples are drawn from the exact FCI wavefunction. Two measurement strategies are compared. First, measurements performed solely in Hartree--Fock basis with the Hartree--Fock configuration in each of the $M$ NOCI measurement bases (``HF ($M=\cdots$)'', dashed lines). Second, measurements are distributed evenly across the Hartree--Fock and $M$ NOCI bases (``NOCI ($M=\cdots$)'' solid lines). (a) Final ground-state energy error as a function of diagonalization subspace size $d$. Increasing $M$ reduces the starting energy error as both measurement strategies benefit from increased classical information. For $M>0$, the energy error varies only weakly with $d$ at small diagonalization sizes. As $d$ increases, the FCI HF curves approach the $M=0$ result. For $M=30$ and $45$, distributing measurements across the NOCI bases yields a modest reduction in energy error at small $d$.
    (b) Zoomed view of panel (a).
    }
    \label{fig:NOCI_Ms}
\end{figure*}

To overcome these sampling inefficiencies, we now investigate whether measurements in a non-orthogonal basis can be used to efficiently generate unique electronic configurations. These non-orthogonal measurements can be combined with terminal orbital rotations typically used in SQD ansazte states like LUCJ and therefore require no increase in circuit depth. By `efficient', we mean both increasing the number of unique configurations from a fixed number of quantum measurements, and that these unique configurations yield a smaller dimensional subspace $d$ for diagonalization without increasing ground-state energies. The first criterion concerns the measurement cost required to obtain unique configurations. Uniform random sampling and noise excel according to this metric, rapidly producing large numbers of distinct bitstrings from a fixed number of measurements. The second criterion requires that these bitstrings correspond to energy-lowering configurations. 

As shown in \cref{fig:draftcartoon}(right), instead of measuring a quantum state directly in the computational basis, we apply one of a set of unitaries $\{U_m\}_{m=1}^{M}$ prior to measurement, thereby sampling from a collection of non-orthogonal configurations. In step (I), unitaries are obtained using a classical NOCI procedure described in \cref{sec:methods}. This classical optimization uses an ansatz $\sum_{m=0}^M c_m U_m|\Psi_0\rangle$ (\cref{eq:noci_opt}) to obtain optimized orbital rotations $U_m$ applied to the Hartree--Fock state $|\Psi_0\rangle$, where $M$ is a tunable hyperparameter.  Measurement outcomes from each of the $M$ orbital bases are restored to the correct particle number using configuration recovery (step (II) in \cref{fig:draftcartoon}(right)), and any spin product expansion is performed (step (III) in \cref{fig:draftcartoon}). Subsequently, in step (IV), a dense molecular Hamiltonian is constructed using measurements in each of the $M$ bases before being diagonalized using the iterative Davidson method \cite{Davidson1975Jan}, which we have generalised to accommodate a non-orthogonal basis. Finally, analogously to standard SQD, average occupancies are computed to inform subsequent rounds of configuration recovery in step (V), but these are now specific to each measurement basis. 

We first consider the impact of the number of NOCI measurement bases in \cref{fig:NOCI_Ms}, where we benchmark strongly correlated electrons in a stretched hydrogen chain. To isolate the effect of the measurement strategy from any initial state preparation, we sample quantum measurements from the Born probabilities of the exact ground state vector (`FCI' state). We additionally ensure that the classical information used to construct the NOCI basis benefits both measurement protocols. For \textit{both} measurement strategies, the diagonalization procedure is given the single Hartree--Fock configuration in each of the the $M$ non-orthogonal bases as well as the computational (Hartree--Fock) basis. In the Hartree--Fock measurement strategy (dashed lines), measurements add additional configurations only to the Hartree--Fock basis. Meanwhile in the NOCI measurement strategy (solid lines), measurements are distributed uniformly across the Hartree--Fock and the $M$ non-orthogonal measurement bases. As a result, both strategies have identical energies at $d=M+1$, where only the initial HF configuration in each basis yield the subspace to diagonalize the Hamiltonian and no measured configurations have yet been added. 

Indeed, \cref{fig:NOCI_Ms}(a), increasing $M$ lowers the initial $d=M+1$ energy for both measurement strategies (Hartree--Fock or NOCI basis), reflecting the additional classical information encoded in the optimized NOCI basis rotations. For $M>0$, the final energy error is nearly independent of diagonalization size at small $d$, indicating that the sampled configurations largely overlap with information already present in the NOCI basis. Distributing measurements across the NOCI bases yields a modest reduction in energy error relative to measuring exclusively in the Hartree--Fock basis, although this effect is only visible for the larger values of $M$. As $d$ increases, the Hartree--Fock-only measurement strategy eventually recovers the same energetic advantage through additional sampled configurations. Consequently, the advantage of distributing measurements across the NOCI bases diminishes as $d$ increases. More significantly, \cref{fig:NOCI_Ms} identifies a regime of practical interest where NOCI measurements are most effective: small diagonalization subspaces and sufficiently large NOCI bases. In this regime, where only a small fraction of the full CI space is retained, measurements distributed across NOCI bases consistently achieve lower energy errors than Hartree--Fock-only measurements. 

\begin{figure*}[!t]
    \centering
    \includegraphics[width=1\linewidth]{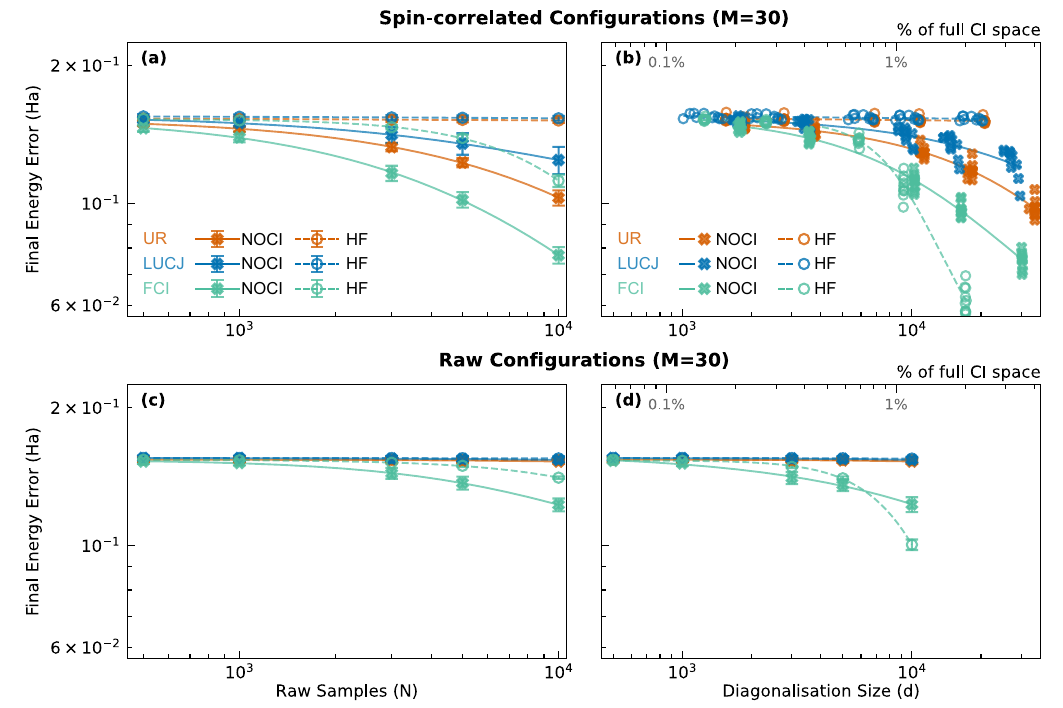}
    \caption{
    SQD for $\mathrm{H}_{12}$ in the STO-6G basis with no frozen orbitals, $R=2.0\AA$, and $M=30$ NOCI measurement bases. Samples are drawn from the exact FCI wavefunction, a CCSD-optimized LUCJ ansatz, or uniformly random configurations (UR). For each sampling method, measurements are performed either exclusively in the Hartree--Fock basis (HF) or distributed uniformly across the Hartree--Fock and NOCI bases (NOCI).
    (a) Results using a restricted spin product expansion, with diagonalization subspaces constructed from all sampled configurations. The $x$-axis denotes the total number of measurements. Each sampled pair $(\alpha_i,\beta_i)$ is expanded to $\{(\alpha_i,\beta_i),(\beta_i,\alpha_i),(\alpha_i,\alpha_i),(\beta_i,\beta_i)\}$. 
    (b) Same as panel (a), but with a fixed diagonalization size $d$; the $x$-axis denotes the number of configurations retained for diagonalization. 
    (c) Raw configurations without spin product expansion, with all sampled configurations included in the diagonalization subspace. 
    (d) Same as panel (c), but with a fixed diagonalization size $d$.
    }
    \label{fig:NOCI}
\end{figure*}

Moving away from sampling the exact FCI wavefunction, we benchmark NOCI and Hartree--Fock methods on quantum states typical of experimental demonstrations in \cref{fig:NOCI} by fixing $M=30$. The idealized performance in sampling the exact FCI state is compared to sampling the CCSD-optimized LUCJ ansatz accommodating hardware constraints, as well as uniform random (UR) sampling to upper-bound the effect of noise in finding new unique configurations. For consistency with previous SQD benchmarks, UR sampling strategies use five rounds of configuration recovery and diagonalization, with only average orbital occupancies propagated between iterations, independently within each measurement basis. The top row uses a restricted spin-correlated expansion in which a sampled bitstring $x=x_\alpha x_\beta$ is expanded to the four configurations $x_\alpha x_\alpha$, $x_\alpha x_\beta$, $x_\beta x_\alpha$, and $x_\beta x_\beta$. Unlike the Cartesian product used previously, this construction preserves correlations between sampled $\alpha$- and $\beta$-spin strings and limits the growth of the diagonalization subspace. 

When combined with the restricted spin-correlated expansion, distributing measurements across the NOCI bases improves sample efficiency for all three sampling strategies in \cref{fig:NOCI}(a). For a fixed measurement budget, NOCI measurements consistently achieve lower energy errors than measurements performed exclusively in the Hartree--Fock basis, with the resulting energies improving beyond the NOCI ground-state energy itself. In contrast when only the raw measured configurations are retained without restricted spin-correlated expansion (\cref{fig:NOCI}(c)), neither LUCJ nor UR sampling substantially improves upon the NOCI ground-state energy over the range of measurement counts considered.

By fixing $M=30$, NOCI is anticipated to outperform HF only in the intermediary regime, $d \in [1000, 6000]$ from our previous results in \cref{fig:NOCI_Ms}. Controlling for diagonalization subspace size $d$, we see this favourable regime in \cref{fig:NOCI}(d) when sampling for the FCI state (green markers). For all other strategies, LUCJ and UR sampling benefit significantly from measurements distributed across the NOCI bases, yielding lower final energy errors than the corresponding Hartree--Fock-only measurement strategy. We find that in direct contrast to the results of \cref{fig:good}, uniform sampling (orange solid) outperforms LUCJ (blue solid) when measured in the NOCI basis in all regimes, even when controlling for diagonalization size $d$, confirming limitations of LUCJ as a universally applicable SQD ans\"atze for molecules, particularly those with correlated or stretched geometries.

These results demonstrate that measurements performed in non-orthogonal NOCI bases can substantially improve the quality of the subspaces constructed for SQD. While NOCI measurements increase the diversity of sampled configurations, the improvements observed after controlling for diagonalization size show that their benefit extends beyond simply generating more unique bitstrings. For both LUCJ and uniform-random sampling, configurations obtained from the NOCI bases enable significantly lower energy errors than those obtained from Hartree--Fock measurements alone. Since practical applications do not have access to exact CI wavefunctions, these results suggest that tailoring the measurement basis is a promising route for both enhancing quantum ans\"atze design and benchmarking noisy sampling for SQD methods.

\section{Discussion \label{sec:discussion}}

Our results highlight an important caveat in the interpretation of existing SQD benchmarks. In standard implementations of SQD, the computational cost of the classical post-processing stage is often only indirectly constrained through quantities such as the total number of measurements or the number of sampled configurations used during subsampling. However, these quantities do not uniquely determine the size of the final diagonalization problem. The number of unique state vectors ultimately used to project the Hamiltonian can grow significantly beyond the number of configurations directly obtained from measurement. This growth in diagonalization subspace size, $d$, occurs through specific mechanisms discussed in \cref{sec:methods} including configuration recovery, configuration carryover, and spin product expansion. Consequently, apparent improvements in SQD performance can originate from an enlarged classical subspace rather than from improved quantum state preparation or more informative sampling distributions.

An important observation is that configuration recovery on noisy samples acts as a stochastic mechanism for generating new configurations. Noise itself increases the rate at which new unique configurations are sampled. During configuration recovery, noisy measurements are probabilistically `corrected' to physically acceptable configurations that can subsequently be incorporated into the diagonalization subspace. When combined with configuration carryover and spin product expansion, these corrected configurations polynomially increase the additional number of unique basis vectors used in diagonalization. In this sense, noise appears to compensate for the sampling inefficiencies discussed by Reinholdt \emph{et al.} \cite{Reinholdt2025Jul} when preparing and evolving a quantum ansatz state noiselessly. Viewed through this lens, the strong performance of noisy SQD in some benchmark regimes becomes less surprising. 

At the same time, care should be taken when extrapolating these observations to larger active spaces. For the molecules considered in this work, the resulting diagonalization subspaces can become a non-negligible fraction of the relevant Hilbert space, making it comparatively easy for stochastic procedures to discover useful determinants. As system size increases, however, the fraction of Hilbert space covered by any fixed resource budget decreases exponentially, and we therefore do not expect configurations produced from configuration recovery to remain competitive with physically informed quantum sampling strategies. Our results should instead be interpreted as highlighting the importance of controlling the final diagonalization dimension when comparing different SQD protocols.

A related consideration applies to the Cartesian-product spin expansion. For small benchmark systems, this procedure can significantly lower the computed energy by generating large numbers of additional configurations. However, the energy-lowering benefit of this procedure is expected to diminish with system size as the generated configuration set becomes an exponentially smaller fraction of the full configuration space. However, even in regimes where the energetic benefit becomes small, the classical overhead remains, further motivating benchmarking methodologies that explicitly report and control the final diagonalization dimension.

Our results demonstrate that choosing non-orthogonal measurement bases provide a more chemically informed route towards generating useful configurations. Measurements distributed across optimized NOCI bases consistently improved sample efficiency relative to measurements performed exclusively in the Hartree--Fock basis, particularly when paired with the restricted spin product expansion implemented in this work. This advantage persisted even after explicitly controlling the final diagonalization dimension for both LUCJ and uniform-random sampling, indicating a higher quality of subspace was discovered. 

We also show that uniform sampling in the NOCI basis provides an important benchmark to guide SQD applications design - comparing uniform random sampling and LUCJ in the NOCI basis in \cref{fig:NOCI} shows the limitations of LUCJ as a quantum ansatz, which is not evident from measurements only in the Hartree--Fock basis in \cref{fig:good}.

\begin{table}[t] 

    \makebox[\linewidth][l]{\textbf{(a) Runtime}}
    
    \vspace{0.3em} 
    \begin{tabular*}{\linewidth}{@{\extracolsep{\fill}}lcc}
        Resource & SQD & NOCI-SQD \\
        \hline 
        \noalign{\vskip 2pt}
        \begin{tabular}[c]{@{}l@{}} 
            State-preparation and\\ basis construction 
        \end{tabular}
        & $O(\kappa^6)$ & $O(\kappa^6+M^2N_{\mathrm{cdf}}\kappa^3)$ \\ 
        Single matrix element & $O(1)-O(N_e^2)$ & $O(N_{\mathrm{cdf}}\kappa^3)$ \\ 
        \begin{tabular}[c]{@{}l@{}} 
            Davidson matrix-\\ 
            vector product 
        \end{tabular} 
        & $O(d^2)$ & $O(d^2)$ \\
        \noalign{\vskip 2pt}
    \end{tabular*}
    
    \vspace{.3em} 
    \makebox[\linewidth][l]{\textbf{(b) Memory}} 
    \vspace{0.3em} 
    \begin{tabular*}{\linewidth}{@{\extracolsep{\fill}}lcc} 
        Resource & SQD & NOCI-SQD \\ 
        \hline 
        \noalign{\vskip 2pt}
        Matrix storage & $O(d^2)$ & $2O(d^2)$ \\ 
        \begin{tabular}[c]{@{}l@{}} 
            Projected matrix\\ 
            structure 
        \end{tabular} 
        & Typically sparse & Generally dense \\ 
    \end{tabular*}

    \caption{Comparison of dominant classical runtime and memory costs between standard SQD and NOCI-SQD. ``State-preparation and basis construction" denotes the classical procedures used to create and optimise an LUCJ ansatz with a CCSD initalization and construct the measurement bases prior to sampling. ``Single matrix element" refers to the evaluation of one projected Hamiltonian matrix element during the construction of the diagonalization problem, while ``matrix storage" denotes the memory required to store the projected matrices used by the Davidson solver. Here $d$ denotes the dimension of the projected diagonalization subspace, $\kappa$ the number of spatial orbitals, $M$ the number of NOCI measurement bases, $N_e$ the number of electrons and $N_{\mathrm{cdf}}$ the number of compressed double-factorization terms. (a) Dominant runtime costs. (b) Dominant memory costs.}
    \label{table:cost_comparison}
\end{table}

Finally, the present NOCI results should be interpreted within the context of the benchmark system considered. Our non-orthogonal measurement protocol was demonstrated on $\mathrm{H}_{12}$, a strongly correlated system in which NOCI methods are expected to provide greater benefit than in weakly correlated molecules such as unstretched $N_2$. Furthermore, the improved sample efficiency comes at the cost of increased classical preprocessing overhead. 

As summarized in \cref{table:cost_comparison}, NOCI-SQD requires construction of an optimized NOCI measurement basis and explicit evaluation of non-orthogonal overlap and Hamiltonian matrix elements between determinants sampled from different orbital bases. Relative to standard SQD, this introduces additional costs associated with NOCI basis construction, transition-density based matrix-element evaluation, and storage of both Hamiltonian and overlap matrices. A detailed derivation of the runtime and memory scalings reported in \cref{table:cost_comparison} is provided in \cref{app:classicalcosts}. In particular, NOCI basis construction scales approximately as $O(M^2N_{\mathrm{cdf}}\kappa^3)$, while evaluation of a single non-orthogonal Hamiltonian matrix element scales approximately as $O(N_{\mathrm{cdf}}\kappa^3)$ due to the compressed double-factorized contractions. By contrast, standard SQD employs Slater--Condon matrix-element evaluations between orthogonal determinants and requires storage of only a single projected Hamiltonian matrix. Consequently, improving the scalability of NOCI basis construction and inter-basis matrix-element evaluation remains an important direction for future work.

\section{Conclusion \label{sec:conclusion}}

Motivated by concerns surrounding sampling inefficiencies in quantum sampling methods such as SQD, we explore whether noisy SQD has regimes where its performance is comparable to classically post-processing uniformly random samples. In this work, we first showed that noisy quantum sampling may lead to an uncontrolled growth of classical resources required for subsequent diagonalization. Under these conditions, replacing quantum measurements with uniform random samples can reproduce or even outperform idealized measurements of quantum states typically used for benchmarking within the literature. These results suggest that the apparent robustness of noisy SQD can originate from an increase in classical resource usage rather than from improvements in the quality of the sampled quantum state. More broadly, our findings establish the diagonalisation subspace size as a critical resource metric that must be explicitly controlled when benchmarking SQD methods.

After controlling diagonalization size, we then address the problem of discovering unique, energy-lowering configurations by introducing a measurement strategy based on non-orthogonal configuration interaction (NOCI). By distributing measurements across bases optimized with respect to the molecular Hamiltonian, we obtained improved sample efficiency relative to measurements performed solely in the Hartree--Fock basis. We confirm these resource improvements persist even after controlling for classical resources required for diagonalization. The regime in which NOCI measurements are most effective corresponds to small diagonalization subspaces, where only a limited fraction of the full CI space can be explored. In this setting, basis engineering provides a chemically motivated mechanism for increasing the quality of sampled configurations.

Perhaps most notably, measurements performed in the NOCI basis remained beneficial even when coupled to poor sampling distributions. Relative to measuring only in the Hartree--Fock basis, our NOCI approach shows promising reductions in ground state energy estimation when sampling from the exact ground state; an approximate ground state, or with uniform random sampling. However, SQD using uniform-sampling in the NOCI basis outperforms all other approaches other than sampling the true ground state in the same basis. This suggests that while our methodology does not entirely overcome limitations with poor quantum ansatz design, SQD with uniform sampling in the NOCI basis could be an additional benchmark for SQD experiments for correlated systems. 

Finally, additional classical overhead is associated with constructing and diagonalizing non-orthogonal subspaces. These costs grow polynomially with system size and arise from NOCI basis construction and the evaluation of inter-basis overlap and Hamiltonian matrix elements. Future work should therefore focus on developing scalable procedures for generating and distributing NOCI measurement bases, reducing the cost of evaluating inter-basis matrix elements, and testing whether the observed benefits persist for larger active spaces and a broader range of weakly and strongly correlated molecular systems.

\section{Methods \label{sec:methods}}

All methods in this paper seek to diagonalize the standard molecular Hamiltonian, restated in second quantization, 
\begin{align}
    H = \sum_{pq} h_{pq} \, a_{p}^{\dagger} a_{q} + \frac12 \sum_{pqrs} h_{pqrs} \, a_{p}^{\dagger} a_{q}^{\dagger} a_{s} a_{r}, \label{eqn:H}
\end{align} where $h_{pq}$ and $h_{pqrs}$ are the one- and two-electron integrals, respectively, and $ a_{p}, a_{q}^{\dagger}$ are fermionic annihilation and creation operators for creating or annihilating an electron in spin-orbital $\varphi_{p}$. For a molecular system of $N_e$ electrons, an approximate wavefunction is obtained by fixing a finite dimensional space of single electrons, 
\begin{align}
    \mathcal{V} := \mathrm{span}\{\phi_\mu\}_{\mu=1}^\kappa,
\end{align}
where $\phi_\mu$ are known basis functions and $\kappa$ is the total number of spin orbitals. In this notation, a single Slater determinant constructed from a set of occupied spin orbitals $\{\varphi_i\}_{i=1}^{N_e}\subset\mathcal V$ may be written as $ a^\dagger_{1} a^\dagger_{2} \cdots a^\dagger_{N_e} \ket{\mathrm{vac}}$. However, each occupied spin orbital is additionally expanded in the basis $\{\phi_\mu\}_{\mu=1}^{\kappa}$ as 
\begin{align} 
    \ket{\varphi_i} = \sum_{\mu=1}^{\kappa} C_{\mu i} \ket{\phi_\mu}, \qquad i=1,\ldots,N_e, 
    \label{eq:occ_orb_basis}
\end{align}
where $C\in\mathbb R^{\kappa\times N_e}$ is the occupied orbital coefficient matrix. The matrix $C$ completely specifies the Slater determinant. Meanwhile, the overlap of basis functions is given by the overlap matrix $S$,
\begin{align}
    S_{\nu \mu} := \langle \phi_\nu \ket{\phi_\mu}.
    \label{eqn:basis-overlap}
\end{align} 
It is typical to normalize the occupation matrix $C$ such that it satisfies
\begin{align}
    C^T S C = \mathbb{I}_{N_e},
\end{align} 
where $S$ is the $\kappa \times \kappa$ basis-function overlap matrix defined in \cref{eqn:basis-overlap} and $\mathbb{I}_{N_e}$ is an $N_e$ dimensional identity matrix \cite{Kjaergaard2026May}.

\subsection{SQD}
SQD is a hybrid quantum–classical method to compute ground-state energies and approximate wavefunctions of electronic structure Hamiltonians. The procedure is summarized in \cref{fig:draftcartoon} (left). Rather than diagonalizing the full Hamiltonian directly, SQD prepares an approximate quantum state $\ket{\Psi}$ assumed to have non-zero overlap with the true ground state $\ket{\Psi_G}$ such that $|\langle \Psi_G | \Psi \rangle|^2 > \epsilon$ for some $\epsilon > 0$. The current SQD literature largely focuses on LUCJ ans\"atze initialized using a restricted closed-shell CCSD calculation. These mappings are well established and further details can be found in Refs.~\cite{Matsuzawa2020Feb,Motta2023Oct}.

The candidate quantum state, $\ket{\Psi}$ is prepared on $n$ qubits, where each qubit corresponds to one of $n/2$ spatial orbitals with either spin up ($\alpha$) and down ($\beta$) state under the Jordan-Wigner mapping. Measurements are performed in the computational (Hartree--Fock) basis, producing a set of bitstrings $\chi'$. Measurement outcomes $x' \in \chi'$ specify the occupations of each spin-orbital. In the noiseless case, sample $x$ is drawn according to the Born probability $|\langle x | \Psi \rangle|^2$, such that configurations with larger amplitudes occur more frequently.

In the presence of noise, measured bitstrings may correspond to nonphysical configurations that violate particle-number or spin constraints. To address this, SQD employs an iterative configuration-recovery procedure \cite{Robledo-Moreno2025Jun}. Given a noisy bitstring $x'$, bits are probabilistically flipped according to the average orbital occupancies estimated from the previous iteration, producing a recovered set of configurations $\chi^{(j)}$ in each iteration $j$. Because bits are corrected individually, configuration recovery can generate physically valid configurations that were not directly measured on the quantum device. 

The recovered configurations are then randomly subsampled into sets $\mathcal{S}_k^{(j)} \subset \chi^{(j)}$ of size $r_0$, where $k=1, \hdots, K$ is the $k$-th batch subset. Additional configurations may then be appended through carryover from previous iterations and subsequently expanded through spin-product expansion and spin symmetrization - details of these procedures are discussed below. Since these procedures can generate configurations that are not contained in $\chi^{(j)}$, we denote the final set used for diagonalization by $\hat{\mathcal{S}}_k^{(j)}$.

Then one proceeds by projecting $H$ to reduced, subspaces $\{\hat{\mathcal{S}}_k\}$,
\begin{align}
    H^{(k)} := P_k H P_k, \quad P_k = \sum_{x \in \hat{\mathcal{S}}_k}\ket{x}\bra{x}, \label{eqn:Hprojection}
\end{align} and diagonalizing $H^{(k)}$ through classical methods (e.g. iterative Davidson \cite{Davidson1975Jan}). The SQD estimate is taken as the lowest eigenvalue obtained among the independently constructed projected subspaces, where eigenvalues satisfy
\begin{align}
    H^{(k)} \mathbf{c} = E \mathbf{c}, 
\end{align} 
where $\mathbf{c}$ is the eigenvector.

This process is repeated for $j$ rounds or until the estimated average occupancies and energies from $\{ \hat{\mathcal{S}}_k^{(j)}\}$ converge to a desired tolerance level, which is set to $10^{-9}$ in our work. We now discuss some important levers in this protocol: 

\textbf{Carryover}: In the standard SQD implementation, configurations whose estimated amplitudes exceed a threshold (default \(10^{-4}\)) are retained from round \(j-1\) and appended to the newly generated random subsets $\{ \mathcal{S}_k^{(j)}\}$. Carrying over unique samples from previous iterations can lead to an uncontrolled growth in the dimension of $\mathcal{S}_k^{(j)}$. Non-unity carryover tends to disproportionately benefit SQD experiments with noise, as demonstrated by supporting simulations in \cref{app:ExtraFigs} (\cref{fig:carryover_app}).

For benchmarking, we set the carryover threshold to unity (i.e. no-carryover) for this paper, including \cref{fig:good,fig:badresult}. Classically tractable benchmarks have small enough active spaces for which uncontrolled growth in the size of $|\mathcal{S}_k^{(j)}|$ will eventually span a large proportion of the full Hilbert space. We also set K=1 for a single batch as a worst case for variance. For each data point, we use ten trials where we independently take $N$ samples, perform configuration recovery, subsample, and solve the eigenvalue problem. 

\textbf{Spin product expansion}: Spin product expansion broadly refers to how recovered bitstrings $x \in \mathcal{S}_k^{(j)}$ are used to further expand the subspace $\hat{\mathcal S}_k^{(j)}$. One splits $x$ into halfstrings $x_\alpha$, $x_\beta$ corresponding to spin up/down configurations, and recombines them using an elected method. One such method forms the Cartesian product of all possible combinations of $\alpha$ and $\beta$-half-strings, resulting in $|\mathcal{S}_k^{(j)}|^2$ configurations. This Cartesian product is the default setting in qiskit-sqd-addon, which cannot be altered, and contributes to the rapid growth in unique samples in noisy SQD as shown in \cref{fig:badresult}. In addition to spin-product expansion, \emph{spin symmetrization} is an optional preprocessing step in which both \(\{x_\alpha\}\) and \(\{x_\beta\}\) are replaced by \(\{x_\alpha\} \cup \{x_\beta\}\) before applying the spin-expansion procedure.

When using these methods, uniform random sampling has the highest rate of growth of $|\hat{\mathcal S}_k^{(j)}|$ and is seen to outperform noiseless LUCJ states for $N_2$ as well as $H_8$ in 6-31g basis, and $H_{12}$ in the STO-6G basis (see \cref{app:ExtraFigs}).

For benchmarking in \cref{fig:good}, \cref{fig:NOCI_Ms} and \cref{fig:NOCI}(c)-(d), we turn off all spin product expansion and spin symmetrization. We introduce a restricted form of spin product expansion in \cref{fig:NOCI}(a)-(b). All other details of the original SQD procedure can be found in Ref.~\cite{Robledo-Moreno2025Jun}.

\subsection{NOCI}

To improve the efficiency with which SQD discovers useful electronic configurations, we consider measurements performed in a collection of non-orthogonal orbital bases rather than exclusively in the Hartree--Fock basis. 

In \cref{fig:draftcartoon}(I), a classical optimization procedure is used to discover orbital rotations $\{U_m\}_{m=0}^{M}$ that define our NOCI measurement basis. To construct this basis, the optimization procedure iterates between optimizing orbital rotations $U_m$ and solving a generalized eigenvalue equation to find the lowest energy eigenvalues using the state, 
\begin{align}
    \sum_{m=0}^{M} c_m U_m |\Psi_0\rangle = \sum_{m=0}^{M} c_m |\Phi_m\rangle.  \label{eq:noci_opt}
\end{align} Here, each $m$-th non-orthogonal determinant $|\Phi_m\rangle := U_m |\Psi_0\rangle$ is defined by applying $U_m$ to the Hartree--Fock reference determinant $|\Psi_0\rangle$. The coefficients $\{c_m\}$ and orbital rotations $\{U_m\}$ are optimized classically to minimize the energy, and $M$ is a tunable hyperparameter for the optimization protocol. The resulting orbital rotations define our measurement bases. Since the NOCI optimization explicitly searches for rotations that improve the description of the ground state, measurements in the corresponding bases preferentially sample configurations expected to be energy-lowering. Additional details concerning the construction and optimization of the NOCI basis are provided in \cref{app:noci_basis}. 

To reduce the computational cost associated with repeated matrix-element evaluation, all NOCI calculations employ a compressed double-factorized (CDF) representation of the molecular Hamiltonian. The electron-repulsion tensor is first approximated using a pivoted Cholesky decomposition and subsequently compressed into a collection of low-rank leaf and core tensors. This representation is used both during NOCI basis construction and during the evaluation of sampled-determinant matrix elements. A detailed description of the CDF procedure is provided in \cref{app:cdf}. 

For a given set of optimized orbital rotations, measurements are performed independently in each basis in \cref{fig:draftcartoon}(II). Samples originating from different measurement bases generally correspond to non-orthogonal determinants and therefore cannot be combined using the standard orthogonal projection procedure of \cref{eqn:Hprojection}. Instead, overlap and Hamiltonian matrix elements are evaluated between sampled determinants belonging to different bases and assembled into a generalized eigenvalue problem for each batch, 
\begin{align}
    \tilde{\mathbf H} ^{(k)}\mathbf c = E \tilde{\Gamma} ^{(k)} \mathbf c, 
\end{align}
where $\tilde{\Gamma}^{(k)}$ is the sampled-determinant overlap matrix and $\tilde{\mathbf H}^{(k)}$ is the corresponding sampled-determinant Hamiltonian matrix for the $k$-th batch. The construction of these matrices and the generalized Davidson procedure used to solve the resulting non-orthogonal diagonalization problem are described in \cref{app:noci_diag}, corresponding to \cref{fig:draftcartoon}(IV).\\

\begin{acknowledgments}
C.v. and S.S are supported by the ARC Centre of Excellence for Engineered Quantum Systems (CE17010000). C.v. and R.S.G. are also supported by UQ’s Queensland Digital Health Center via funding from UQ’s Health Research Accelerator (HERA) initiative. R.S.G is supported by access to Pawsey Supercomputing Research Center via the 
National Computational Merit Allocation Scheme 2026. C.v. is further supported by the University of Queensland's Graduate School, and the Queensland Government Department of Environment, Science and Innovation. All authors acknowledge useful conversations and feedback within the Quantum Working Group for Health Care and Life Sciences.
\end{acknowledgments}

\subsection*{Author contributions} 
C.v. performed the SQD benchmarking studies, developed and integrated the software framework, carried out the simulations, analyzed the data, and led manuscript preparation. J.C. conceived the NOCI measurement-basis approach and developed the associated software implementations. C.v. and J.C. jointly developed and integrated the NOCI-SQD workflow. J.C. and R.S.G. contributed to the interpretation of results and provided ongoing technical feedback throughout the project. S.S. co-supervised the project and provided feedback on the analysis and manuscript. R.S.G. conceived the original research direction and co-supervised the project. C.v., J.C., and R.S.G. co-wrote the manuscript. All authors discussed the results, reviewed the manuscript, and approved the final version.

\subsection*{Competing interests} 
All authors declare no financial or non-financial competing interests.
\subsection*{Data and code availability statement} 
All scripts and data are available upon reasonable request and will be made publicly available at time of publishing.
\subsection*{AI usage statement} The authors used ChatGPT (GPT-5.5, OpenAI; https://chatgpt.com) to supplement standard tools for literature search. C.v.R and J.C used Claude Sonnet 5.0 to draft efficient runtime or memory implementations of existing functions, to draft code documentation, and enable data visualization. All scientific content, analysis, and conclusions is the creative output of the authors -- AI tools were not used in writing, editing or analysis of main text or appendices.

\bibliography{apssamp}

\clearpage
\appendix
\crefalias{section}{appendix} 
\crefalias{subsection}{appendix}
\crefalias{subsubsection}{appendix}
\section{Notation and background \label{app:background}}

Consistent notation used in the main text and all appendices is summarized in \cref{tab:notation}. We present key implementation details for constructing the NOCI orbital rotations, where we use compressed double factorization to efficiently store and contract matrices with the molecular Hamiltonian. 

\begin{table*}[!t]
    \centering
    \footnotesize
    \begin{tabular}{l|l}
     Symbol & Definition \\
    \hline
        \multicolumn{2}{c}{\textit{Operators and matrices}} \\
    \hline
       $a_p ~(a^\dagger_q)$ & Annihilation (creation) operators for spin orbitals $p$ ($q$) \\
       $H$ & Molecular electronic structure Hamiltonian \\
       $H_{\mathrm{cdf}}$ & Compressed double-factorized representation of $H$ \\
       $\mathbf H_{mn}$ & NOCI Hamiltonian matrix element, $\langle\Psi_0|U_m^\dagger H_{\mathrm{cdf}}U_n|\Psi_0\rangle$ \\
       $h_{pq}$ & One electron integrals \\
       $h_{pqrs}$ & Two electron integrals (electron-repulsion integral tensor) \\
       $L_{pq}^{(t)}$ & $t$-th Cholesky factor \\
       $P_k$ & Projector onto subspace $\hat{\mathcal{S}}_k$. \\
       $S_{\mu\nu}$ & Overlap matrix of known basis functions $\bra{\phi_\mu} \phi_\nu \rangle $\\
       $U_m$ & Unitary rotation for the $m$-th measurement basis. $U_0:=I$ for the Hartree Fock computational basis\\
       $U^{t}$ & Orthogonal matrix diagonalizing $L^{(t)}$ \\
       $Z_{kl}^{t}$ & Core tensor associated with CDF factor $t$ \\
       $\Gamma_{mn}$ & NOCI overlap matrix, $\langle\Psi_0|U_m^\dagger U_n|\Psi_0\rangle$ \\
       $\gamma^{(mn)}$ & Transition one-particle density matrix between $\ket{\Phi_m}$ and $\ket{\Phi_n}$ \\ 
       $\rho^\sigma$ & Spin sector one-particle reduced density matrix constructed from the generalized eigenvector coefficients $\{c_i\}$ \\
    \hline
        \multicolumn{2}{c}{\textit{States, wavefunctions and vectors}} \\
    \hline
       $\mathbf c$ & Coefficient vector in generalized eigenvalue problems \\
       $C$ & Occupied-orbital coefficient matrix defining a Slater determinant \\
       $\varphi(\cdot)$ & Single particle spin orbital wavefunction  \\
       $\phi(\cdot)$ & Known basis function for $\varphi(\cdot)$ \\
       $\Psi(\cdot)$ & $N_e$ multi-particle wavefunction with all spatial and spin degrees of freedom for each particle \\
       $\ket{\Psi}$ & Prepared quantum state sampled by SQD \\
       $\ket{\Psi_G}$ & Reference ground state obtained from FCI or CASCI \\
       $\ket{\Psi_0}$ & Hartree--Fock reference determinant \\
       $\ket{\Phi_m}$ & Rotated determinant, $\ket{\Phi_m}=U_m\ket{\Psi_0}$ \\
    \hline
        \multicolumn{2}{c}{\textit{Dimensionality constants and scalars}} \\
    \hline
       $E$ & Eigenvalue of a generalized eigenvalue problem \\
       $E_{\mathrm{loc}}^{(mn)}$ & Local energy associated with determinant pair $(m,n)$ \\
       $d$ & The dimension of the diagonalization subspace, $d = |\hat{\mathcal{S}}_k^{(j)}|$ \\
       $o_{m,k}^{\sigma}$ & Average occupation of orbital $\mu$ in measurement basis $U_m$ and spin sector $\sigma$ \\
       $n$ & Number of qubits \\
       $N_e$ & Number of particles/electrons \\
       $N$ & Number of total quantum measurements or samples\\
       $N_{\mathrm{cho}}$ & Number of Cholesky vectors \\
       $N_{\mathrm{cdf}}$ & Number of retained compressed double-factorization terms \\
       $R$ & Molecule bond length\\
       $r_j$ & The number of elements per batch for all $k$ i.e. $r_j:= |\mathcal{S}_k^{(j)}|$. For no carryover, $r_j = r_0 \forall j$ \\
       $\kappa$ & Total number of spin orbitals. Equivalently, the number of qubits\\
       $\lambda$  & Local depolarizing noise rate \\
    \hline
        \multicolumn{2}{c}{\textit{Indices}} \\
    \hline
       $j=1, \hdots, J$ & Iteration index $j$ for a total of $J$ iterations in SQD \\
       $k=1, \hdots, K$ & Batch index $k$ for a total of $K$ batches in SQD \\
       $m=0, \hdots, M$ & Measurement index $m$ for a total of $M+1$ measurement bases, where $M=0$ is the Hartree Fock basis\\
       $\mu = 1, \hdots, \kappa$ & An index over known functions $\phi(\cdot)$\\
       $\alpha$ ($\beta$) & The left (right) half-strings of $x$ or $x'$ corresponding to orbitals with up (down) electron spin \\
       $\sigma$ & $\sigma \in \{\alpha, \beta\}$ \\
    \hline
        \multicolumn{2}{c}{\textit{Sampled set notation}} \\
    \hline
       $x' \in \mathcal{\chi}'$ & Total set $\mathcal{\chi}'$ of sampled configurations $x'$ from a (noisy) preparation of $\ket{\Psi}$ i.e. $N= |\mathcal{\chi}'|$ \\
       $x \in \mathcal{\chi}^{(j)}$ & Total set, $\mathcal{\chi}$, of noiseless ($ \mathcal{\chi}'\equiv \mathcal{\chi}^{(j)} \forall j$) or noisy recovered configurations $x$ at the $j$-th iteration\\
       $\mathcal{S}_k^{(j)}$ & A randomly sampled subset $\mathcal{S}_k^{(j)} \subset \mathcal{\chi}^{(j)}$ for $k$-th batch in the $j$-th iteration \\
       $\hat{\mathcal{S}}_k^{(j)}$ & \begin{tabular}[c]{@{}l@{}} 
            Final configurations in $k$-th batch of the $j$-th iteration used for diagonalization from $\mathcal{S}_k^{(j)}$ after\\ spin product expansion/symmetrization
        \end{tabular}
        \\
       $\mathcal{\chi}_m$ & Set of configurations measured in basis $U_m$ \\
       $\tilde{\mathcal S}_m$ & Subsampled determinant set from basis $m$ \\ 
       $\tilde{\Gamma}$ & Sampled-determinant overlap matrix \\ 
       $\tilde{\mathbf H}$ & Sampled-determinant Hamiltonian matrix \\ 
       $\tilde{\gamma}$ & Sampled-determinant transition density matrix \\
       $\tilde{E}_{\mathrm{loc}}$ & Sampled-determinant local energy \\
    \end{tabular}
    \caption{Summary of notation used throughout the main text and appendices. }
    \label{tab:notation}
\end{table*}

\section{Supporting analysis of SQD in Hartree--Fock basis}
\label{app:ExtraFigs}

We support our discussion of classical mechanisms leading to the uncontrolled growth of configurations in the main text using additional simulations. 

\begin{figure}[!t]
    \centering
    \includegraphics[width=1.\linewidth]{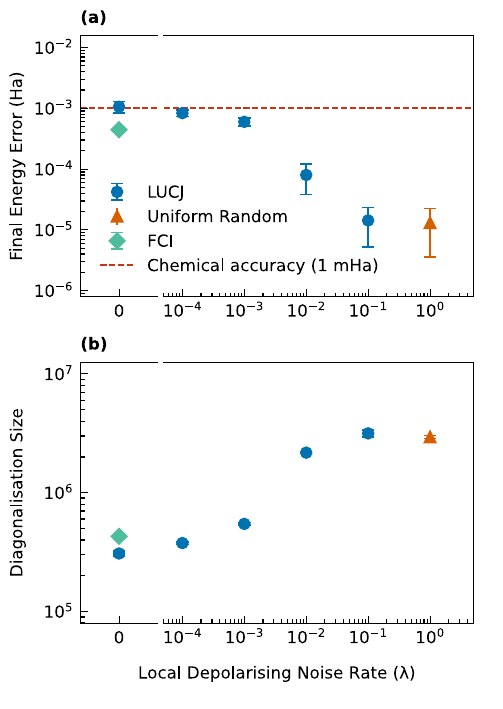}
    \caption{
    Noisy SQD for $\mathrm{H}_8$ in the 6-31g basis with zero frozen orbitals and bond length $R=2.0\AA$. In all panels, the carryover threshold is unity so that growth in $d$ is due to spin-product expansion of (noisy) samples. (a) Final ground-state energy error for standard SQD under increasing local depolarizing noise rates applied to all two-qubit gates in the LUCJ circuit. The LUCJ ansatz (circles) is initialized via compressed double-factorization of CCSD $t_2$ amplitudes (LUCJ-DF) and optimized to minimize reconstruction error. This is compared to sampling from the exact FCI wavefunction (“FCI”, diamonds) or uniformly randomly from all configurations (triangles). Each data point uses $r_j = 9{,}650$ unique configurations for up to five iterations, where only average orbital occupancies are propagated between iterations for configuration recovery. 
    (b) Growth of the diagonalization size $d$ for data in panel (a) due to Cartesian products of $\alpha$- and $\beta$-spin configurations on (noisy) measurements, i.e. underlying sample diversity increases from left to right.
    }
    \label{fig:badH8_app}
\end{figure}

\begin{figure*}[!t]
    \centering\includegraphics[width=1\linewidth]{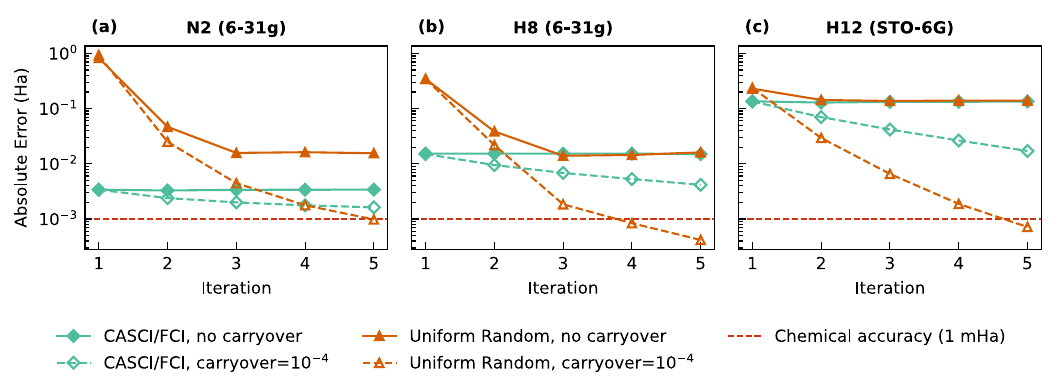}
    \caption{
    Effect of the SQD carryover threshold on ground-state energy estimation starting with 1000 configurations, $r_0 = 1000$. (a) $\mathrm{N}_2$ in the 6-31g basis with two frozen orbitals, $R=1.0\AA$. (b) $\mathrm{H}_8$ in the 6-31g basis, $R=2.0\AA$. (c) $\mathrm{H}_{12}$ in the STO-6G basis, $R=2.0\AA$. Results are shown for sampling from the exact CASCI($\mathrm{N}_2$)/FCI($\mathrm{H}$-chains) wavefunction (diamonds) and from uniformly random configurations (triangles). Solid lines correspond to a carryover threshold of unity, such that no configurations are retained between SQD iterations, while dashed lines correspond to the default SQD carryover threshold of $10^{-4}$. 
    }
    \label{fig:carryover_app}
\end{figure*}

Our first simulation shows the strong performance of noisy SQD may be dominated by the classical generation of additional configurations rather than improved sampling efficiency of a quantum state under SQD. We choose a strongly correlated system, $\mathrm{H}_8$, instead of uncorrelated $N_2$ system in the main text. Sampling from the true FCI state should be less concentrated for highly correlated molecules $\mathrm{H}_8$ (supported on $3\times10^{-3}$\% full CI space) compared to uncorrelated $\mathrm{N}_2$ (supported on $1\times10^{-4}$\% full CI space). Moreover, the strong correlation present is not expected to be captured adequately by the CCSD-initialized LUCJ ansatz. Under this picture, introducing noise should further reduce the probability of sampling physically relevant configurations. Despite this, the behavior observed for $\mathrm{H}_8$ closely mirrors that of $\mathrm{N}_2$ discussed in \cref{fig:badresult}. In \cref{fig:badH8_app}(a), increasing depolarizing noise lowers the final SQD energy error despite degrading the quality of the underlying quantum state. Uniform random (UR) sampling and LUCJ circuits with high local depolarizing noise rates both substantially outperform noiseless LUCJ and even exact FCI sampling when the diagonalization size is allowed to grow without constraint. As in \cref{fig:badresult}, these improvements correlate strongly with the growth of the diagonalization space shown in \cref{fig:badH8_app}(b), indicating that the dominant effect is the classical generation of additional configurations rather than improved overlap with the true ground state wavefunction.

While the main text focuses on the growth of the diagonalization subspace arising from spin-product constructions, carryover between SQD iterations provides a second mechanism by which the effective search space over physically sensible configurations can expand. Because carryover is enabled by default in many SQD implementations, it is useful to understand how carryover contributions interact with noise and spin-product expansion discussed in the main text.  Namely, we show here that carryover rapidly expands the diagonalization size for SQD with noisy configurations compared to SQD with noiseless quantum sampling. 

\Cref{fig:carryover_app} illustrates the impact of carryover on benchmark performance by comparing procedures with and without the propagation of configurations between iterations. For CASCI/FCI sampling (diamonds), disabling carryover by setting the threshold to unity results in iteration-independent behavior, with only small fluctuations arising from the stochastic subsampling procedure. Introducing the default carryover threshold of $10^{-4}$ produces a gradual reduction in energy error over successive iterations as configurations identified in earlier rounds accumulate within the diagonalization subspace. The effect is considerably more pronounced for uniform random sampling (triangles). Without carryover, the energy error decreases rapidly during the first few iterations before plateauing once the average orbital occupancies used by configuration recovery have converged, typically after three iterations. Enabling carryover transforms this plateau into a persistent downward trend, allowing the diagonalization subspace to continue growing as useful configurations discovered in previous iterations are retained. These results further support the interpretation that iterative SQD performance can be strongly influenced by the continual expansion of the effective diagonalization space, independent of the quality of the underlying sampling distribution.

\section{Construction of the NOCI Measurement Basis \label{app:noci_basis}}  
The non-orthogonal measurement bases used in this work are generated using a non-orthogonal configuration interaction (NOCI) ansatz constructed from a Hartree--Fock reference determinant $|\Psi_0\rangle$ and a set of orbital rotations, 
\begin{align} 
    |\Psi_M\rangle &= \sum_{m=0}^{M} c_m U_m |\Psi_0\rangle, \nonumber \\
    &= \sum_{m=0}^{M} c_m |\Phi_m\rangle
\end{align} 
where $U_0:=I$, $|\Phi_m\rangle :=U_m |\Psi_0\rangle$ and, for $m>0$,
\begin{align}
    U_m=\exp\left(\sum_{pq}\theta^{(m)}_{pq}a_p^\dagger a_q\right).
\end{align}

The orbital rotations are optimized variationally using the compressed Hamiltonian introduced in \cref{app:cdf}. For a fixed set of rotations, the NOCI wavefunction is obtained by solving the generalized eigenvalue problem 
\begin{align} 
    \mathbf H \mathbf{c} = E \Gamma \mathbf{c}, \label{eqn:generaleigevalue_app} 
\end{align} 
where 
\begin{align} 
    \Gamma_{mn} &= \langle\Psi_0| U_m^\dagger U_n |\Psi_0\rangle, \\ 
    \mathbf H_{mn} &= \langle\Psi_0| U_m^\dagger H_{\rm cdf} U_n |\Psi_0\rangle = \Gamma_{mn} E_{\mathrm{loc}}^{(mn)}, 
\end{align} 
with 
\begin{align} 
    E_{\mathrm{loc}}^{(mn)} = &2h_{pq}\gamma_{pq}^{(mn)} +2h_{pqrs}^{\mathrm{cdf}} \gamma_{qp}^{(mn)} \gamma_{rs}^{(mn)} \nonumber \\
    &-h_{pqrs}^{\mathrm{cdf}} \gamma_{qr}^{(mn)} \gamma_{ps}^{(mn)}, 
\end{align} 
and 
\begin{align} 
    \gamma^{(mn)} := C_n \left( C_m^T S C_n \right)^{-1} C_m^T 
\end{align} 
is the transition one-particle density matrix between determinants $|\Phi_m\rangle$ and $|\Phi_n\rangle$. Here $C_m$ denotes the occupied orbital coefficient matrix of the determinant $|\Phi_m\rangle$, as introduced in \cref{eq:occ_orb_basis}. Likewise, $S$ is the basis-function overlap matrix defined in \cref{eqn:basis-overlap}.

The basis is constructed iteratively through alternating determinant selection and orbital optimization stages. During determinant selection, the NOCI basis is constructed iteratively, beginning from the Hartree--Fock determinant. Suppose the first $m-1$ determinants have already been selected and the generalized eigenvalue problem, \cref{eqn:generaleigevalue_app}, has been solved in the corresponding subspace. Let
\begin{align}
    |\Psi_{m-1}\rangle
    =
    \sum_{n=0}^{m-1}
    c_nU_n|\Psi_0\rangle
\end{align}
denote the resulting NOCI wavefunction and let $E_0$ be its lowest eigenvalue. The next orbital rotation is obtained by maximising 
\begin{align} 
    F(U_m) = \frac{|\langle \Phi_m | (H_\text{cdf}-E_0) |\Psi_{m-1}\rangle|^2} {\max(E_{\mathrm{diag}}(U_m)-E_0,\delta)}, \label{eqn:greedy_objective}
\end{align}

where $\delta$ is a small regularization parameter and $E_{\mathrm{diag}}(U_m)$ is defined as,
\begin{align} 
    E_{\mathrm{diag}}(U_m) := \frac{ \langle \Phi_m|H_\text{cdf}|\Phi_m\rangle }{ \langle \Phi_m|\Phi_m\rangle }.
\end{align}
Following basis expansion, all orbital rotations are refined through global optimization of the NOCI Rayleigh quotient,
\begin{align}
    \min E:=  \frac{ \bra{\Psi_m} H \ket{\Psi_m} }{\langle\Psi_m \ket{\Psi_m}}, \quad \langle \Psi_m \ket{\Psi_m} = 1.\label{eqn:Var_app}
\end{align} 

The final optimized rotations $\{U_m\}_{m=0}^{M}$ define the measurement bases used throughout the SQD calculations. 

\section{Compressed Double-Factorized Hamiltonian \label{app:cdf} } 
To reduce the cost of evaluating Hamiltonian matrix elements, all NOCI calculations in this work employ the compressed double-factorization (CDF) representation of the molecular Hamiltonian \cite{Cohn2021Dec,Oumarou2024Jun}. The four-index tensor $h_{pqrs}$, appearing in \cref{eqn:H}, is reshaped into a two-index $pq \times rs$ Coulomb matrix $h_{(pq)(rs)}$. Since this matrix is positive semi-definite, one may use a pivoted Cholesky decomposition \cite{PivotedCholesky2022Nov} to reduce computational resources and factorise,
\begin{align}
    h_{(pq)(rs)} \approx \sum_{t=1}^{N_\text{cho}} L_{pq}^{(t)} L_{rs}^{(t)}.
\end{align} 
While storage of the original tensor is $\mathcal{O}(\kappa^4)$, the matrix $L$ is of size $\mathcal{O}(\kappa^2 N_\text{cho})$ \cite{Krisiloff2015Nov}. 
For each Cholesky leaf $L^{(t)}$, we diagonalize
\begin{align}
    L^{(t)} = U^{t}\Lambda^{t}(U^{t})^T,
\end{align}
where $\Lambda^t=\mathrm{diag}(\lambda^t_k)$. Defining
\begin{align}
    Z^t_{kl} := \lambda^t_k\lambda^t_l,
\end{align}
one obtains
\begin{align}
    h_{(pq)(rs)} &\approx \sum_{t}^{N_\text{cdf}} \sum_{k,l=1}^\kappa U^t_{pk}U^t_{qk} Z^t_{kl}U^t_{rl}U^t_{sl} =: H_\text{cdf},
    \label{eqn:appendix_cdf}
\end{align}
Here the sum over $t$ is given by user-specified parameter $N_\text{cdf} \leq N_\text{cho}$. In our work we set $N_\text{cdf} = N_\text{cho}$, where a leaf and core tensor is computed for each Cholesky leaf, $L_{pq}^{(t)} L_{rs}^{(t)}$. 

This compressed double factorized representation of the Cholesky tensor can optionally be further optimised by minimizing the reconstruction error of the electron-repulsion integral tensor subject to orthogonality and symmetry constraints \cite{Oumarou2024Jun,Cohn2021Dec},
\begin{align}
    \sum_{k=1}^\kappa U^t_{pk}U^t_{qk} = \delta_{pq},& 
    \quad
    \sum_{k=1}^\kappa U^t_{kp}U^t_{kq} = \delta_{pq} \quad \forall t \\
    Z^t_{kl} &= Z^t_{lk}\quad \forall t.
\end{align} 

Computational details for compressed double factorization including optimization loss function, constraints and numerical approaches, are found in Ref.~\cite{Cohn2021Dec}. 

\section{Non-Orthogonal Diagonalization of Sampled Determinants \label{app:noci_diag}}

\begin{figure*}[!t]
    \centering
    \includegraphics[width=1\linewidth]{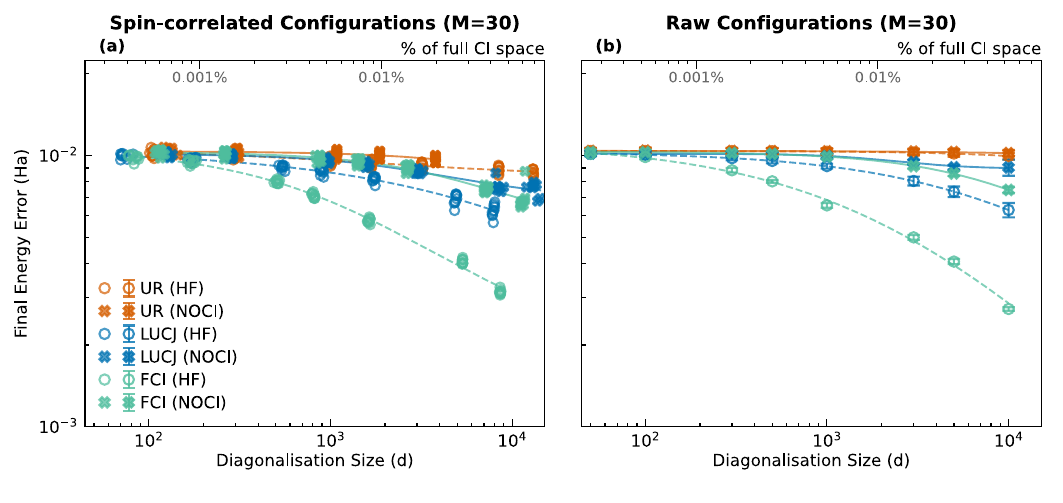}
    \caption{
    SQD for $\mathrm{N}_{2}$ in the 6-31g basis with two frozen orbitals, $R=1.0\AA$, and $M=30$ NOCI measurement bases. Samples are drawn from the exact FCI wavefunction, a CCSD-optimized LUCJ ansatz, or uniformly random configurations (UR). For each sampling method, measurements are performed either exclusively in the Hartree--Fock basis (HF) or distributed uniformly across the Hartree--Fock and NOCI bases (NOCI).
    (a) Results using a restricted spin product expansion. The $x$-axis denotes the number of configurations retained for diagonalization after spin product expansion. Each sampled pair $(\alpha_i,\beta_i)$ is expanded to $\{(\alpha_i,\beta_i),(\beta_i,\alpha_i),(\alpha_i,\alpha_i),(\beta_i,\beta_i)\}$. 
    (b) Raw configurations without spin product expansion. The $x$-axis denotes the number of configurations retained for diagonalization, using sampled configurations only. 
    }
    \label{fig:N2NOCIapp}
\end{figure*}

Let $\mathcal{\chi}_m$ denote the set of determinants sampled after measurement in basis $U_m$. Determinants originating from different measurement bases are generally non-orthogonal and therefore cannot be combined using the standard SQD projection procedure. Instead, for each basis $m$, we construct random subsets $\tilde{\mathcal{S}}_m \subset \mathcal{\chi}_m$ obtained from a frequency-based subsampling procedure across all bases. For determinants $x_i\in\tilde{\mathcal{S}}_m$ and $x_j\in\tilde{\mathcal{S}}_n$, we define overlap matrix elements
\begin{align} 
    \tilde{\Gamma}_{ij} = \langle x_i | x_j \rangle, 
\end{align}
and 
\begin{align} 
    \tilde{\mathbf H}_{ij} = \langle x_i | H_{\mathrm{cdf}} | x_j \rangle = \tilde{\Gamma}_{ij} \tilde{E}_{\mathrm{loc}}^{(ij)}. 
\end{align} 
The matrix-element evaluation follows the same determinant-overlap and transition-density formalism used in \cref{eqn:generaleigevalue_app}, but is applied to sampled determinants rather than orbital-rotated reference determinants. To distinguish the two constructions, we use the sampled-determinant overlap $\tilde{\Gamma}$, sampled-determinant transition density $\tilde{\gamma}$, sampled-determinant Hamiltonian matrix $\tilde{\mathbf H}$, and sampled-determinant local energy $\tilde{E}_{\mathrm{loc}}$.
The matrix elements reduce to the familiar orthogonal SQD expressions when $m=n$, 
\begin{align} 
    \tilde{\Gamma}_{ij}=\delta_{ij}, \qquad x_i, x_j \in \tilde{\mathcal{S}}_m, 
\end{align} 
and become non-trivial only for determinants belonging to different orbital-rotation bases. Let $C_i$ and $C_j$ denote the occupied-orbital coefficient matrices associated with sampled determinants $x_i$ and $x_j$, respectively. The overlap matrix between the two determinants factorizes into spin sectors, 
\begin{align} 
    \tilde{\Gamma}_{ij} = \det\!\left[ (C_{i,\alpha})^\dagger C_{j,\alpha} \right] \det\!\left[ (C_{i,\beta})^\dagger C_{j,\beta} \right]. 
\end{align}

Provided the spin-sector overlap matrices are nonsingular, the corresponding transition density matrices are 
\begin{align} 
    \tilde{\gamma}^{\sigma}_{ij} = C_{j,\sigma} \left[ (C_{i,\sigma})^\dagger C_{j,\sigma} \right]^{-1} (C_{i,\sigma})^\dagger . 
\end{align}
and define the spin-summed transition density matrix 
\begin{align} 
    \tilde{\gamma}_{ij} = \tilde{\gamma}_{ij}^{\alpha} + \tilde{\gamma}_{ij}^{\beta}.
\end{align}

The sampled-determinant local energy is evaluated using the same compressed double-factorized Hamiltonian introduced in \cref{app:cdf}. Defining 
\begin{align} 
    \tilde{E}_{\mathrm{loc}}^{(ij)} = E_{\mathrm{nuc}} + \tilde{E}_{1}^{(ij)} + \tilde{E}_{2}^{(ij)}, 
\end{align} 
the one-electron contribution is 
\begin{align} 
    \tilde{E}_{1}^{(ij)} = \sum_{pq} h_{pq} \tilde{\gamma}_{ij;qp}, 
\end{align}

and the two-electron contribution is evaluated through the CDF factors $\{U^t,Z^t\}$. For each factor $t$, we define the rotated transition density matrices 
\begin{align} 
    D_{ij,\sigma}^{t} := (U^t)^T \tilde{\gamma}_{ij,\sigma} U^t, 
    \label{eqn:rot_trans_density_app}
\end{align} 
and the corresponding spin-summed diagonal densities 
\begin{align} 
    d_{ij,k}^{t} := D^{t}_{ij,\alpha;kk} + D^{t}_{ij,\beta;kk}. 
\end{align} 

The two-electron contribution is then 
\begin{align} 
    \tilde{E}_{2}^{(ij)} = \frac12 \sum_{t=1}^{N_{\mathrm{cdf}}} \sum_{kl} Z_{kl}^{t} \Big(& d_{ij,k}^{t} d_{ij,l}^{t} - D_{ij,\alpha;kl}^{t} D_{ij,\alpha;lk}^{t} \nonumber \\
    &- D_{ij,\beta;kl}^{t} D_{ij,\beta;lk}^{t} \Big). 
\end{align} 

We then solve the generalised eigenvalue problem
\begin{align} 
    \tilde{\mathbf H}\mathbf{c} = E\tilde{\Gamma}\mathbf{c}, \label{eqn:noci_sqd_gevp} 
\end{align} 
using a generalized Davidson algorithm. 

At each iteration, the current estimate $\mathbf{c}$ is expressed in a small trial subspace $V$, and the problem is projected onto this subspace to give
\begin{align}
    \tilde{\mathbf H}_V = V^T \tilde{\mathbf H} V, \qquad \tilde\Gamma_V = V^T \tilde\Gamma V .    
\end{align}

The reduced problem $\tilde{\mathbf H}_V \mathbf{y} = E\, \tilde\Gamma_V \mathbf{y}$ is solved by standard diagonalization giving an approximate eigenpair $(E, \mathbf{c})$ with $\mathbf{c} = V\mathbf{y}$.

The residual
\begin{align}
    \mathbf{r} = \tilde{\mathbf H} \mathbf{c} - E\,\tilde\Gamma \mathbf{c}
\end{align}
is used to generate a new trial vector via a correction rule
\begin{align}
    t_i = -\frac{r_i}{\tilde{\mathbf H}_{ii} - E},
\end{align}

which is added to the subspace $V$ and the process repeats. The subspace is periodically restarted from the current best estimate to limit its size. Iteration continues until $\|\mathbf{r}\|_2$ falls below a convergence threshold.

\subsection{Average orbital occupations from the NOCI state} 

In order to apply configuration recovery in the subsequent iteration, we require the average orbital occupancies with respect to each measurement basis. We construct the spin sector one-particle reduced density matrices 
\begin{align} 
    \rho^\sigma_{pq} = \frac{ \sum_{ij} c_i^* c_j \tilde{\Gamma}_{ij} \tilde{\gamma}^\sigma_{ij;pq}} 
    {\sum_{ij} c_i^* c_j \tilde{\Gamma}_{ij}}.
\end{align}

For each measurement basis $U_m$, the one-particle density matrix is rotated according to 
\begin{align} 
    \rho_m^\sigma = U_m^\dagger \rho^\sigma U_m, 
\end{align} 
and the average occupation of orbital $\mu$ in basis $m$ is defined as 
\begin{align} 
    o_{m,\mu}^{\sigma} = \left(\rho_m^\sigma\right)_{\mu\mu}. 
\end{align} 

\subsection{$\mathbf{N}_2$ NOCI Measurement Results}
\label{app:N2NOCI}

The behaviour for $\mathrm{N}_2$ differs markedly from that observed for the strongly correlated $\mathrm{H}_{12}$ system. As shown in \cref{fig:N2NOCIapp}, measurements distributed across the NOCI sectors provide no improvement over measurements performed exclusively in the Hartree--Fock basis when examined through the lens of a fixed diagonalization subspace. Indeed, the best performance is obtained by sampling the exact CASCI wavefunction in the Hartree--Fock basis, followed by LUCJ sampling in the Hartree--Fock basis. In contrast to \cref{fig:NOCI}, neither the restricted spin product expansion nor measurements in the NOCI bases provide an advantage for this weakly correlated molecule. Nevertheless, NOCI measurements continue to alleviate the sampling concentration problem. For example, obtaining a diagonalization subspace of $10^4$ configurations requires on the order of $10^7$ measurements in the Hartree--Fock basis, whereas the same diagonalization size can be reached with fewer than $10^5$ measurements when sampling is distributed across the NOCI bases. Together with the $\mathrm{H}_{12}$ results, this suggests that measurement-basis engineering is effective at increasing configurational diversity, but that the resulting energy benefits are most promising in systems where important electronic configurations are not efficiently represented in the Hartree--Fock basis.

\subsection{Classical runtime and memory scaling for NOCI}
\label{app:classicalcosts}

The dominant classical costs introduced by NOCI-SQD arise during state-preparation and basis construction, evaluation of non-orthogonal matrix elements, and storage of the projected matrices used in the generalized Davidson procedure. 

\paragraph{State-preparation and basis construction.} The complexity of computations performed prior to quantum sampling includes both LUCJ state preparation and construction of the NOCI measurement basis. The LUCJ parameters are initialized from CCSD amplitudes obtained through an iterative coupled-cluster calculation. Conventional CCSD has a computational cost that scales as $O(\kappa^6)$ with system size \cite{purvisFullCoupledclusterSingles1982} and is the bottleneck in state initialization. This contributes a common $O(\kappa^6)$ preprocessing cost to both SQD and NOCI-SQD. 

The additional NOCI basis construction cost arises from solution of the generalized eigenvalue problem defined in \cref{eqn:generaleigevalue_app}. Since the NOCI basis contains $M+1$ determinants, the overlap and Hamiltonian matrices contain $O(M^2)$ determinant pairs. For each pair $(m,n)$, evaluation of the Hamiltonian matrix element requires construction of the transition density matrix $\gamma^{(mn)}$ and the local energy $E_{\mathrm{loc}}^{(mn)}$. The dominant contribution comes from the two-electron term, which is evaluated using the compressed double-factorized Hamiltonian. From \cref{eqn:appendix_cdf}, each CDF factor contains an orbital-rotation matrix $U^t\in\mathbb{R}^{\kappa\times\kappa}$. Evaluation of the rotated transition density matrices therefore involves dense matrix multiplications whose cost scales as $O(\kappa^3)$ per CDF factor. Summing over all $N_{\mathrm{cdf}}$ factors gives an $O(N_{\mathrm{cdf}}\kappa^3)$ cost per determinant pair and therefore an overall basis-construction cost of 
\begin{align}
    O(M^2N_{\mathrm{cdf}}\kappa^3). 
\end{align}

\paragraph{Non-orthogonal matrix-element evaluation.} Following sampling, determinants originating from different measurement bases are generally non-orthogonal. Each overlap and Hamiltonian matrix element therefore requires explicit evaluation of the transition density matrices and the rotated transition density matrices defined in \cref{eqn:rot_trans_density_app}. As above, the dominant cost arises from the dense matrix multiplications involving the $\kappa\times\kappa$ CDF orbital rotations, yielding a runtime of 
\begin{align} 
    O(N_{\mathrm{cdf}}\kappa^3) 
\end{align} 
for a single non-orthogonal Hamiltonian matrix element. 

In contrast, matrix elements in standard SQD are evaluated using Slater--Condon rules. Diagonal matrix elements involve a double sum over occupied orbitals and scale as $O(N_e^2)$. Matrix elements corresponding to single excitations involve a single sum over occupied orbitals and scale as $O(N_e)$. Matrix elements corresponding to double excitations require only a constant number of integral evaluations, having $O(1)$ cost. Finally, matrix elements corresponding to greater than double excitations are zeroed and require no integral evaluation, scaling $O(1)$.

\paragraph{Memory requirements.} For a projected diagonalization subspace of dimension $d$, standard SQD stores a single projected Hamiltonian matrix and therefore requires $O(d^2)$ memory. NOCI-SQD additionally requires storage of the overlap matrix associated with the generalized eigenvalue problem, increasing the memory requirement to approximately $2O(d^2)$. Furthermore, the projected matrices are generally dense due to non-orthogonal inter-basis couplings, whereas the projected Hamiltonian in standard SQD is typically sparse because of the Slater--Condon selection rules.

\end{document}